\documentstyle[aps]{revtex}
\begin{document}
\title{Scattering of energetic particles 
by anisotropic  magnetohydrodynamic
turbulence with a Goldreich-Sridhar power spectrum}
\author{Benjamin D. G. Chandran }
\address{Department of Physics \& Astronomy, 
University of Iowa, Iowa City, IA 52242-1479 \\
benjamin-chandran@uiowa.edu }
\maketitle

\begin{center}
--- \hspace{0.3cm} accepted for publication in {\em Physical Review Letters} 
\hspace{0.3cm} ---
\end{center}

\begin{abstract}
Scattering rates for a Goldreich-Sridhar (GS) spectrum of 
anisotropic, incompressible, magnetohydrodynamic
turbulence are calculated in the quasilinear approximation.
Because the small-scale fluctuations are constrained to have wave
vectors nearly perpendicular to the background magnetic field,
scattering is too weak to provide either  the mean free paths 
commonly used in Galactic cosmic-ray propagation models
or the mean free paths required for acceleration
of cosmic rays at quasi-parallel shocks.
Where strong pitch-angle scattering occurs, it is due to fluctuations not
described by the GS spectrum, such as fluctuations
generated by streaming cosmic rays.

\end{abstract}
\widetext 
The scattering of energetic particles by turbulent magnetic and
electric fields plays an important role in the acceleration and
propagation of cosmic
rays~\cite{bel78,bla78,dru83,kul69,ces80,ber90,jok66}. The turbulent
fields responsible for cosmic-ray scattering can be excited by the
cosmic rays themselves or by some mechanism that is independent of the
cosmic rays. This paper focuses upon the latter case.  In previous
treatments of scattering, different turbulence models have been used,
including fluctuations with wave vectors~${\bf k}$ parallel to the
ambient large-scale magnetic field~${\bf B}_0$ (slab symmetry) or
perpendicular to ${\bf B}_0$ (2D), or power spectra that are isotropic in
$k$-space~\cite{jok66,bie94,sch98,gia99}. On the other hand, a number
of studies suggest that in magnetohydrodynamic (MHD) turbulence
excited by large-scale stirring, small-scale fluctuations have
non-zero values of $k_\parallel$ that are $\ll k_\perp$, where
$k_\perp$ and $k_\parallel$ are the components of ${\bf k}$ $\perp$
and $\parallel$ to ${\bf B}_0$~\cite{gol95,she83,ng}.  In this paper,
the quasilinear approximation~\cite{sti92} is used to calculate
general scattering rates for incompressible MHD turbulence and also
shear-Alfv\'enic turbulence on the non-MHD scales shorter than the
collisional mean free path of thermal particles~\cite{gol95}.  These
rates are then evaluated for the Goldreich-Sridhar power
spectrum~\cite{gol95}, which has significant power at small scales
only for $k_\perp \gg k_\parallel$.  The condition $k_\perp \gg
k_\parallel$ is found to significantly decrease the efficiency of
pitch-angle scattering relative to the slab-symmetric and isotropic
cases.  Astrophysical applications and limitations of quasilinear
theory (QLT) are discussed.

It is assumed that there is an inertial-range spectrum of fluctuations
extending from some large scale~$l$ to a much smaller scale~$d$, with
the fluctuations at scales $\sim l$ dominating the total magnetic
energy. Only cosmic rays with gyroradii $\rho\ll l$ are considered.  A
scale $l^\prime$ is introduced, with $\rho \ll l^\prime \ll l$.  The
energetically dominant fluctuations on scales $> l^\prime$ are treated
as a uniform field~${\bf B}_0$.  The magnetic fluctuations on
scales~$< l^\prime$, denoted ${\bf B}_1$, are small compared to ${\bf
B}_0$ and are treated using QLT.  It can be verified {\em a
posteriori} that the QLT scattering rates are independent of
$l^\prime$ to lowest order in $\rho/l$.  In contrast to most previous
treatments, the turbulence is treated as strong, in the sense that
fluctuations decorrelate in one linear wave period.

In QLT, the turbulence causes the cosmic
rays to diffuse in momentum space, with the diffusion coefficients
determined by the statistical properties of the 
turbulence~\cite{sch93},
\begin{equation} 
\frac{\partial f}{\partial t} = \frac{\partial}{\partial \xi}
\left( D_{\xi \xi} \frac{\partial f}{\partial \xi}\right) + 
\frac{\partial}{\partial \xi}\left(D_{\xi p}\frac{\partial f}{\partial p}\right)
+ \frac{1}{p^2}
\frac{\partial}{\partial p}\left[ p^2 \left(
D_{p \xi}\frac{\partial f}{\partial \xi}
+ D_{pp}\frac{\partial f}{\partial p}\right)\right],
\label{eq:pad} 
\end{equation} 
where $f$ is the cosmic-ray distribution function averaged over the
small scales of the fluctuating fields, ${\bf p} $ is 
momentum, $\theta$ (the pitch angle) is the angle between ${\bf p}$
and ${\bf B}_0$, and $\xi = \cos \theta$. In equation~(\ref{eq:pad})
it has been assumed that the length scale characterizing
variations in $f$ is large compared to $\rho$, so that $f$ can be
taken to be independent of gyrophase.
At each ${\bf k}$, ${\bf B}_1$ and the incompressible turbulent velocity ${\bf U}_1$
are decomposed into shear-Alfv\'en and pseudo-Alfv\'en components by 
projecting along the appropriate polarization vectors~\cite{gol95}.
 These components are
denoted respectively by the superscripts s and p, so that
\begin{equation} 
{\bf B}_1({\bf k} ,t)   =  {\bf B}_1^{\rm s}({\bf k} ,t) 
+ {\bf B}_1^{\rm p}({\bf k} ,t) ,
\end{equation} 
with an analogous equation for $
{\bf U}_1({\bf k} ,t) $.
The electric field is given by Ohm's Law,
${\bf E}_1 = - (1/c){\bf U}_1 \times {\bf B}_0$.
The normalized power spectra of the
shear-Alfv\'en modes are given by
\begin{eqnarray} 
M^{\rm s}(k_\perp, k_\parallel, \tau) & 
= & \langle {\bf B}^{\rm s}_1({\bf k} , t) \cdot 
{\bf B}_1^{\rm s\, \ast}({\bf k},t+\tau) \rangle/B_0^2 , \\
C^{\rm s}(k_\perp, k_\parallel, \tau) & 
= & \langle {\bf U}^{\rm s}_1({\bf k} , t) \cdot 
{\bf B}_1^{\rm s\, \ast}({\bf k},t+\tau) \rangle/v_A B_0 , 
\mbox{ \hspace{0.3cm} and} \\
K^{\rm s}(k_\perp, k_\parallel, \tau) & 
= & \langle {\bf U}^{\rm s}_1({\bf k} , t) \cdot 
{\bf U}_1^{\rm s\, \ast}({\bf k},t+\tau) \rangle/v_A^2 , 
\end{eqnarray} 
with analogous equations for the pseudo-Alfv\'en modes, where
$v_A$ is the Alfv\'en speed associated
with ${\bf B}_0$, and $\langle \dots \rangle$ denotes an ensemble
average. It is assumed that the turbulence is homogeneous and
stationary, that $\langle {\bf B}_1({\bf x},t) {\bf B}_1({\bf x}+{\bf
r},t+\tau) \rangle = \langle {\bf B}_1({\bf x},t) {\bf B}_1({\bf x}-{\bf
r},t+\tau) \rangle$ with analogous equations for
$\langle {\bf U}_1 {\bf U}_1\rangle$ and $\langle {\bf U}_1 {\bf B}_1
\rangle$
(no magnetic or kinetic helicity), that
$\langle {\bf U}_1({\bf x},t) {\bf B}_1({\bf
x}+{\bf r},t+\tau) \rangle = \langle {\bf U}_1({\bf x},t) {\bf B}_1({\bf
x}+{\bf r},t-\tau) \rangle$ (which gives $D_{\xi p} = D_{p \xi}$), 
and that the shear-Alfv\'en and
pseudo-Alfv\'en modes are statistically independent. 

The contributions to the momentum diffusion coefficients from
the shear-Alfv\'en modes and pseudo-Alfv\'en modes are, respectively,
\begin{equation}
\left(\begin{array}{c}
D^{\rm s}_{\xi \xi} \\ D^{\rm s}_{\xi p} \\ D^{\rm s}_{pp} 
\end{array} \right) =
\lim_{L\rightarrow \infty}\Omega^2 (1-\xi^2)\int \frac{d^3k}{L^3}\, \int_0^{\infty} d\tau
\sum_{n=-\infty}^{\infty}
e^{-i(k_\parallel v_\parallel + n\Omega)\tau}
\frac{n^2 J_n^2(z)}{z^2}
\left(
\begin{array}{c}
\displaystyle
\delta^2 \xi^2 K^{\rm s} + 2 \delta \xi C^{\rm s}
+ M^{\rm s} \\
\displaystyle
(\delta^2 \xi K^{\rm s} + \delta C^{\rm s})(-p) \\
\displaystyle
p^2 \delta^2 K^{\rm s}
\end{array}
\right), \mbox{ \hspace{0.1cm} and}
\label{eq:sagen} 
\end{equation}
\begin{equation}
\left(\begin{array}{c}
D^{\rm p}_{\xi \xi} \\ D^{\rm p}_{\xi p} \\ D^{\rm p}_{pp} 
\end{array} \right) =
\lim_{L\rightarrow \infty}\Omega^2 (1-\xi^2)\int \frac{d^3k}{L^3}\, \int_0^{\infty} d\tau
\sum_{n=-\infty}^{\infty}
e^{-i(k_\parallel v_\parallel + n\Omega)\tau}
\frac{ k_\parallel^2 J_n^{\prime\,2}(z)}{k^2}
\left(
\begin{array}{c}
\displaystyle
\delta^2 \xi^2 K^{\rm p} + 2 \delta \xi C^{\rm p}
+ M^{\rm p} \\
\displaystyle
(\delta^2 \xi K^{\rm p} + \delta C^{\rm p})(-p) \\
\displaystyle
p^2 \delta^2 K^{\rm p}
\end{array}
\right),
\label{eq:pagen} 
\end{equation}
where $\delta = v_A/v$, $z=k_\perp\rho$, $ \rho= v_\perp/\Omega$,
$\Omega$ is the cosmic-ray gyrofrequency, $v_\perp$ and $v_\parallel$
are the cosmic-ray velocity components $\perp$ and $\parallel$ to
${\bf B}_0$, $L$ is the dimension of a window function that multiplies
the variables before a Fourier transform is taken, and the arguments
of $ K$, $ C$, and $ M$ are $({\bf k}, \tau)$.  Since the
shear-Alfv\'en and pseudo-Alfv\'en modes are statistically
independent, $D_{\xi \xi} = D_{\xi \xi}^{\rm s} + D_{\xi \xi}^{\rm
p}$, etc. Equations~(\ref{eq:sagen}) and (\ref{eq:pagen}) are derived
using a standard method based on the linearized Vlasov equation~\cite{sti92},
modified to treat strongly turbulent
fluctuations instead of waves satisfying linear dispersion relations.
Alternatively, they can be derived from equations (7a), (7b), and (7c)
of~\cite{sch93}, if one notes the typographical
error on the eighth line of equation (7a), namely, that
$Q_{R\parallel}$ should instead be $Q_{\parallel R}$.

A Goldreich-Sridhar spectrum of strong, anisotropic MHD 
turbulence~\cite{gol95} is now assumed, with
\begin{equation}
M^{\rm s}(k_\perp, k_\parallel, \tau)
= 
\frac{L^3 }{6\pi}k_\perp^{-10/3} l^{-1/3} g\left(\frac{k_\parallel}{
k_\perp^{2/3}l^{-1/3}}\right) 
e^{-|\tau|/\tau_{\bf k}}
\label{eq:Ms} 
\end{equation} 
for $(l^\prime)^{-1} < k_\perp < d^{-1}$ with $d\rightarrow 0$, where
$\tau_{\bf k} = (l/v_A)(k_\perp l)^{-2/3}$ is the Lagrangian
correlation time appropriate for strong anisotropic incompressible MHD
turbulence, and
\begin{equation}
g(x) = \left\{ \begin{array}{ll}
1 & \mbox{ \hspace{0.3cm} if $|x| < 1$} \\
0 & \mbox{ \hspace{0.3cm} if $|x| \geq 1$} \\
\end{array} \right. .
\label{eq:g} 
\end{equation} 
The spectrum of equation~(\ref{eq:Ms}) is also taken to
describe the fluctuations on scales between $l^\prime$ and $l$,
and the normalization has been chosen so that the total
magnetic energy $  \int _{l^{-1}}^{\infty} k_\perp d k_\perp
\int_{-\infty}^{\infty} d k_\parallel M^{\rm s}(k_\perp, k_\parallel,
0)B_0^2/4 = L^{3} B_0^2/8\pi$.
At small scales in MHD turbulence, there is equipartition between
magnetic and kinetic energies, so that $K^{\rm p} =
M^{\rm p}$ and $K^{\rm s} =M^{\rm s}$.
It is assumed that $M^{\rm p} =
M^{\rm s}$ and $C^{\rm p} = C^{\rm s} = \sigma M^{\rm s}$,
where the arguments of each of these spectra are $(k_\perp,
k_\parallel,\tau)$, and where the fractional cross helicity~$\sigma \in
(-1, 1)$ is independent of ${\bf k}$.  

In many applications, there are two small parameters,
\begin{equation} 
\epsilon = \frac{v}{l\Omega}, \mbox{ \hspace{0.3cm} and \hspace{0.3cm} } 
\delta = \frac{v_A}{v}.
\end{equation}
For $\sin \theta \gg \epsilon^{1/2}$, one finds from 
equations~(\ref{eq:sagen}) and (\ref{eq:pagen}) and the
assumed forms of the power spectra that
to lowest order in $\epsilon$ and $\delta$,
\begin{equation} 
\left(\begin{array}{c}
D^{\rm s}_{\xi \xi} \\ D^{\rm s}_{\xi p} \\ D^{\rm s}_{pp} 
\end{array} \right) =
 \frac{v}{l}\left[
\frac{2 \epsilon^{3/2}|\cos\theta|^{11/2}}{13 \sin \theta}
\sum_{n=1}^{n=\infty} n^{-9/2} - \frac{\delta  \ln \epsilon }{3} \sin^2\theta\right]
\left(\begin{array}{c}
1 \\ -\sigma p \delta \\ p^2 \delta^2
\end{array} \right), \mbox{ and}
\label{eq:sagen2} 
\end{equation} 
\begin{equation} 
\left(\begin{array}{c}
D^{\rm p}_{\xi \xi} \\ D^{\rm p}_{\xi p} \\ D^{\rm p}_{pp} 
\end{array} \right) =
\frac{v}{l}\left\{
\frac{2 \epsilon^{3/2}|\cos\theta|^{7/2} \sin\theta}{13 }
\sum_{n=1}^{n=\infty} n^{-9/2} - \frac{\delta  \ln \epsilon}{6}
 \frac{ \sin^4 \theta }{\cos^2\theta}
\left[1 - \frac{v_A}{v_\parallel} \arctan \left(
\frac{v_\parallel}{v_A}\right)\right]\right\}
\left(\begin{array}{c}
1 \\ -\sigma p \delta \\ p^2 \delta^2
\end{array} \right).
\label{eq:pagen2} 
\end{equation} 
The terms on the right-hand sides of equations~(\ref{eq:sagen2}) and
(\ref{eq:pagen2}) proportional to $\epsilon^{3/2}$ correspond to
fluctuations satisfying the magnetostatic gyroresonance condition
$k_\parallel v_\parallel = n\Omega$, which states that the
Doppler-shifted frequency of a static magnetic fluctuation in the
reference frame of an energetic particle's motion along ${\bf B}_0$ is
an integral multiple of the particle's gyrofrequency.  A
fluctuation is seen as static when a cosmic ray passes through one
wavelength of the fluctuation in a time $(k_\parallel
v_\parallel)^{-1} \ll \tau_{\bf k}$.  The gyroresonant terms in
equations~(\ref{eq:sagen2}) and (\ref{eq:pagen2}) are much smaller
than in the case of slab-symmetric or isotropic turbulence for $\sin
\theta \gg \epsilon^{1/2}$ because equations~(\ref{eq:Ms}) and
(\ref{eq:g}) imply that $k_\perp > k_\parallel^{3/2} l^{1/2}$, so
that fluctuations satisfying $k_\parallel
v_\parallel = n\Omega$ 
 also satisfy $k_\perp \rho > n^{3/2} \epsilon^{-1/2} \sin\theta 
\cos^{-3/2}\theta \gg 1$. The condition $k_\perp \rho \gg 1$ 
implies that a cosmic ray traverses many uncorrelated
fluctuations of the required $k_\parallel$ during a single gyro
orbit. The effects of these uncorrelated fluctuations tend to cancel.
The weakening of gyroresonant scattering due to this gyro-orbit
averaging would occur for any power spectrum in which all fluctuations
on scales $\ll l$ satisfy $k_\perp \gg k_\parallel$.
The terms on the right-hand sides of
equations~(\ref{eq:sagen2}) and (\ref{eq:pagen2}) proportional to
$(-\ln \epsilon) \delta$ correspond to non-resonant interactions. 
In equation~(\ref{eq:pagen2}) the non-resonant term arises from
the $n=0$ term in equation~(\ref{eq:pagen}), which
represents the effects of the magnetic-mirror force of the
pseudo-Alfv\'en modes (transit-time damping). This term
becomes large as $\theta\rightarrow \pi/2$, since as $v_\parallel
\rightarrow v_A$ particles can ``surf'' magnetic mirrors moving at
speeds $\sim v_A$ more effectively.

For $\sin\theta \ll \epsilon^{1/2}$, scattering is dominated by
magnetostatic gyroresonant interactions with shear-Alfv\'en modes, and
to lowest order in $\epsilon$ and $\delta$
\begin{equation}
\left(\begin{array}{c}
D^{\rm s}_{\xi \xi} \\ D^{\rm s}_{\xi p} \\ D^{\rm s}_{pp} 
\end{array} \right) =
\left( \frac{v}{l}\right)  \frac{\pi \theta^2}{8} 
\left(\begin{array}{c}
1 \\ -\sigma p \delta \\ p^2 \delta^2
\end{array} \right).
\label{eq:sagen3} 
\end{equation}
Although $D_{\xi \xi}$ vanishes as $\theta\rightarrow 0$, the 
 pitch-angle scattering frequency~$\nu = 2 D_{\xi
\xi}/(1 - \xi^2)$ (which unlike $D_{\xi \xi}$ is
independent of $\theta$ for isotropic
scattering) approaches $(\pi/4) (v/l)$ as $\theta \rightarrow 0$.
Gyroresonant interactions are stronger for $\theta \lesssim
\epsilon^{1/2}$ than for $\theta \gg \epsilon^{1/2}$, because when
$\theta \lesssim \epsilon^{1/2}$ modes satisfying 
$k_\parallel v_\parallel = n\Omega$ also
satisfy $k_\perp \rho \lesssim 1$, so that a cosmic ray doesn't
traverse many uncorrelated resonant modes during a single gyro orbit.

\begin{figure}[h]
\vspace{8.5cm}
\includegraphics{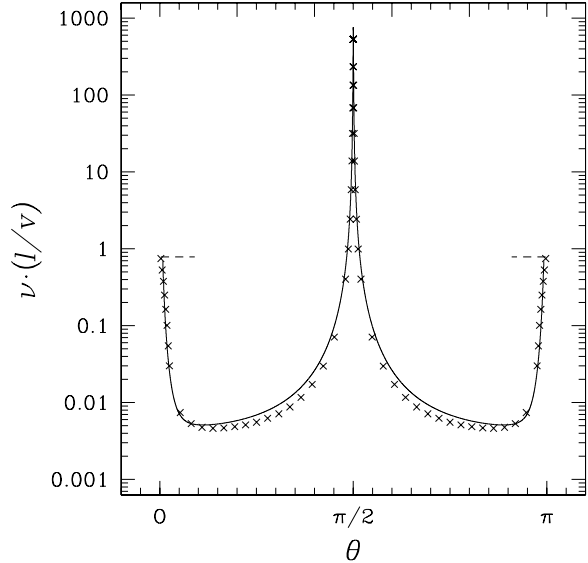}
\caption{The $\theta$ dependence of the pitch-angle scattering
frequency $\nu= 2 (D^{\rm s}_{\xi \xi}+ D^{\rm p}_{\xi \xi})/(1-\xi^2)$
for $\epsilon = \delta = 10^{-3}$. 
The $\times$s indicate
numerical evaluations of equations~(\ref{eq:sagen}) and (\ref{eq:pagen}),
the solid line gives the analytic results of equations~(\ref{eq:sagen2})
and (\ref{eq:pagen2}), and the dashed lines give the limiting value 
from equation~(\ref{eq:sagen3}) of $\nu$ as $\sin \theta\rightarrow 0$.
\label{fig:comp} }
\end{figure}

In figure~\ref{fig:comp}, the pitch-angle scattering frequency~$\nu$
from equations~(\ref{eq:sagen2}) and (\ref{eq:pagen2}) is plotted with
the solid line, and the limiting value of $\nu$ as $\sin\theta
\rightarrow 0$ from equation~(\ref{eq:sagen3}) is given by the dashed
line. The $\times$s indicate numerical evaluations of $\nu$ from
equations~(\ref{eq:sagen}) and (\ref{eq:pagen}) for the
assumed spectra in which only those
terms in the infinite sum with $|n| \leq 10$ are kept and in
which $l^\prime = 0.1 l$ (see introduction). The
values $\epsilon = 10^{-3}$ and $\delta = 10^{-3}$ have been used.
($\sigma$, which only weakly affects $\nu$, has been set to 0.)
The characteristic values $\nu \sim v/l$ for $\sin \theta <
\epsilon^{1/2}$,  $\nu\sim \delta^{-1}
v/l$ for $|\pi - \theta| < \delta$, and
$\nu \sim [(-\ln \epsilon)\delta +
\epsilon^{3/2}]v/l$ for moderate pitch angles can be extrapolated to all values
of $\epsilon$ and $\delta$ much less than~1 in the quasilinear approximation.

When $\nu$ is sufficiently large, $f_0$ becomes
nearly isotropic, and the pitch-angle-averaged distribution
$\overline{f}$ can be treated in the diffusion
approximation~\cite{jok66,sch89},
\begin{equation}
\frac{\partial \overline{f} }{\partial t}   =
\frac{\partial }{\partial l}  \left( \kappa_\parallel
\frac{\partial \overline{f}}{\partial l}  \right)
+ \frac{1}{p^2}\frac{\partial }{\partial p} \left( p^2
\overline{ D}_p \frac{\partial \overline{f} }{\partial p} 
\right) + \dots,
\label{eq:diff} 
\end{equation} 
where $l$ is distance along a field line, and
the ellipsis indicates the omission of the
advection and adiabatic-acceleration terms. 
The coefficient of spatial diffusion along ${\bf B}_0$ is given 
by~\cite{sch89}
\begin{equation} 
\kappa_\parallel    =  \frac{v^2}{8}\int_{-1}^{1} d\xi \frac{(1-\xi^2)^2}{D_{\xi \xi}}.
\label{eq:kpar} 
\end{equation}
To lowest order in $\epsilon$ and $\delta$ when
$\epsilon^{3/2} \ll (-\ln \epsilon)\delta $, QLT gives
\begin{equation}
\kappa_\parallel^{\rm QLT}
= v l (-\delta \ln \epsilon)^{-1}\left(\frac{5}{2} - \frac{3\pi}{4}\right).
\label{eq:kappap} 
\end{equation} 
[In this paper, from equation~(\ref{eq:sagen2}) on,
$M^{\rm p} = M^{\rm s}$; however, if either
$M^{\rm p}$ or $M^{\rm s}$ is set to zero,
equation~(\ref{eq:kappap}) becomes $\kappa_\parallel^{\rm QLT} = vl (-2\delta
\ln \epsilon)^{-1}$.]
To lowest order in $\epsilon$ and $\delta$ when
$( -\ln \epsilon) \delta \ll
\epsilon^{3/2}$ QLT gives
\begin{equation}
\kappa_\parallel ^{\rm QLT}=  vl c_1 (-\delta \ln \epsilon)^{-5/11} \epsilon^{-9/11},
\mbox{ \hspace{0.3cm} where}
\label{eq:kappap2} 
\end{equation} 
$c_1 = (\pi/22) \csc(5\pi/11) 6^{5/11} [(2/13)
\sum_{n=1}^{\infty} n^{-9/2}]^{-6/11}\simeq 0.88$. 
The pitch-angle-averaged momentum diffusion coefficient in 
equation~(\ref{eq:diff}) is given by~\cite{sch89}
\begin{equation} 
\overline{ D}_p  =  \frac{1}{2}\int_{-1}^{1} d\xi \left[ D_{pp} 
- \frac{D_{\xi p}^2}{D_{\xi \xi}}\right].
\label{eq:dp0} 
\end{equation} 
To lowest order in $\epsilon$ and $\delta$,
QLT gives 
\begin{equation}
\overline{D}_p^{\rm QLT}  =
(\pi/24)(1 - \sigma^2)(-\ln \epsilon)
p^2 v_A^2/(vl).
\label{eq:dp} 
\end{equation} 
When $\sigma^2 = 1$, $\overline{ D}_p$ vanishes since
the small-scale fluctuations all travel in a single direction along
${\bf B}_0$ at the speed $v_A$, and, in the reference frame that
follows their motion, particle energies are conserved.

Although QLT is a useful and standard tool, it suffers from important
inaccuracies.  QLT assumes that during the time a particle is
correlated with a turbulent fluctuation, the orbit of that particle is
the same as in a uniform magnetic field.  However,
field-strength fluctuations $\triangle |B|$ with $\triangle |B|/|B|
\equiv \alpha \ll 1$ magnetically trap cosmic rays with $|\xi |
\lesssim \alpha^{1/2}$.  (For incompressible turbulent fluctuations,
which have phase velocities $\sim v_A$ along the magnetic field, and
for cosmic rays with $v \gg
v_A$, the trapping condition is essentially the same as if the
fluctuations were stationary.)  The trajectories of such
trapped particles differ greatly from the trajectories of particles in
a uniform field, violating the QLT assumptions.  Because the integral
in equation~(\ref{eq:dp0}) is dominated by values of $|\xi|\lesssim
\delta\ll 1$ for which trapping is important, the value of $\overline{
D}_p$ in equation~(\ref{eq:dp}) 
is unreliable.  Similarly, when $( -\ln \epsilon) \delta \ll
\epsilon^{3/2}$, the integral in equation~(\ref{eq:kpar}) is dominated
by small $|\xi|$, and thus the value of $\kappa_\parallel$
in equation~(\ref{eq:kappap2}) is unreliable.
In addition, assuming an unperturbed particle orbit in the presence of a
slowly and non-periodically
varying  ${\bf E}_1$ or ${\bf B}_1$ leads to spurious changes in
a particle's magnetic moment $\mu = mv_\perp^2/2B_0$, which, as an
adiabatic invariant, should be virtually conserved when ${\bf E}_1$
and ${\bf B}_1$ vary on a time scale $\gg \Omega^{-1}$.  The
non-resonant terms in equations~(\ref{eq:sagen2}) and (\ref{eq:pagen2})
arise from slowly varying modes and imply such spurious changes in
$\mu$, thereby significantly overestimating non-resonant pitch-angle
scattering.
Since $\kappa_\parallel$ in equation~(\ref{eq:kappap}) is
determined by this non-resonant pitch-angle scattering, equation~(\ref{eq:kappap})
underestimates~$\kappa_\parallel$.

Although the QLT results for the key particle-transport coefficients are
inaccurate, QLT does show that
resonant scattering by MHD turbulence with $k_\perp \gg k_\parallel$
is much weaker than resonant scattering by slab-symmetric or isotropic
fluctuations. Moreover, equation~(\ref{eq:kappap}) as a lower bound on
$\kappa_\parallel$ has important implications.
If $B_0 = 5 \mu$G, $l= 100$ pc, and $v_A = 10^6$ cm/s [parameters
characteristic of the interstellar medium (ISM)], then $\epsilon =
2.2 \times 10^{-9} E_{\rm GeV}$ for a relativistic proton, where
$E_{\rm GeV}$ is the proton's energy in GeV, and $\delta = 3.3 \times
10^{-5}$. If $E_{\rm GeV} \ll 10^6$, then $\epsilon ^{3/2} \ll (-\ln
\epsilon)\delta$, and 
equation~(\ref{eq:kappap}) gives a lower limit to the
scattering mean free path
$\kappa_\parallel/v$ of  $430 \mbox{ kpc} \times (20 - \ln E_{\rm
GeV})^{-1}$.  This value is so large that if the power-law spectrum of
interstellar turbulence inferred from observations~\cite{spa98} is
described by equation~(\ref{eq:Ms}), then some mechanism besides such
turbulence must be invoked to explain the confinement of cosmic rays
to the Galaxy~\cite{ber90}.  At energies $\lesssim 10^2 - 10^3$ GeV,
such a mechanism is provided by resonant MHD waves that cosmic rays
themselves excite, but at higher energies it is believed that
self-confinement does not work~\cite{ces80,ber90}.  For $E_{\rm GeV} >
10 ^2 - 10^3$, cosmic-ray confinement and isotropization can be
explained even if turbulent scattering is weak if one takes into
account molecular-cloud magnetic mirrors~\cite{cha00}.  If the
interstellar turbulence generated by supernovae and stellar
winds is described by equation~(\ref{eq:Ms}), the inefficient
scattering associated with such turbulence may indicate that 
quasi-parallel shocks
are unable to accelerate cosmic rays up to the $\sim 10^6$ GeV
energies at the ``knee'' of the galactic cosmic-ray energy
spectrum~\cite{bla78,lag83}, although this suggestion is
controversial~\cite{ell00}.  Quasi-perpendicular shocks, however, may
be able to accelerate cosmic rays to the knee and
beyond~\cite{jok87,kir96}.

I thank Steve Spangler, Eliot Quataert, Torsten Ensslin, 
Randy Jokipii, Steve Cowley, Jon Arons, Peter Goldreich, and 
Russell Kulsrud for valuable discussions.

\end{document}